\documentstyle[12pt]{article}
\begin{document}
\begin{titlepage}

\begin{flushright}
{\large                                TUIMP-TH-96/77}
\end{flushright}
\vspace{0.3in}
\begin{center}
{\Large\bf Yukawa Corrections to Top Quark Production at the LHC in 
Two-Higgs-Doublet Models}
\vspace{.5in}

{\bf   Hong-Yi Zhou $^{a,b}$, Chong-Sheng Li $^{a,c}$ 
and Yu-Ping Kuang $^{a,b}$}\\
\vspace{.3in}
      $^a$ CCAST (World Laboratory), \hspace{.2cm}
      P. O.\hspace{0.2cm}  Box 8730, Beijing 100080, P.R. China,\\

  $^b$ Institute of Modern Physics and Department of Physics,\\
   Tsinghua University, Beijing 100084, P.R. China $^*$\\

$^c$ Physics Department, Peking University, Beijing 100871, P.R. China$^*$\\

\vspace{.5in}

\end{center}

\begin{footnotesize}
\begin{center}\begin{minipage}{5in}

\begin{center} ABSTRACT\end{center}

~~~The $O(\alpha m_t^2/m_W^2)$ Yukawa corrections to top quark pair 
production by $gg$ fusion at the LHC  are calculated in 
two-Higgs-doublet models. We find that the correction to the 
cross-section can exceed about $-10\%$ for certain parameter values.

\end{minipage}\end{center}
\end{footnotesize}
\vspace{.7in}

~~PACS number: 14.80Dq; 12.38Bx; 14.80.Gt

\vspace{0.5in}

\vfill
{\flushleft \hspace{0.5cm}$^*$ Mailing address }
\end{titlepage}

\eject
\rm
\baselineskip=0.36in

\begin{flushleft} {\Large\bf I. Introduction} \end{flushleft}

Recently, the existance of the top quark is 
confirmed by the CDF and D0 Collaborations with the mass and production
cross-section $m_t=176\pm 8(stat)\pm 10(syst)$ GeV~~~ 
$\sigma=6.8^{+3.6}_{-2.4} pb$,
~and $m_t=199^{+19}_{-20}(stat)\pm 22(syst)$ GeV~~~
$\sigma=6.4\pm 2.2 pb$, respectively \cite{CDFD0} .  
This is a remarkable success of the Standard Model (SM) since
this measured mass is close to the central value predicted 
by the best fit of the SM to the LEP precision electroweak data. 
There are a number of new and interesting issues related to the top quark, 
such as the precision measurement of the mass, width and Yukawa couplings 
through its direct production and subsequent decay at both hadron 
and $e^+e^-$  colliders. At the future multi-TeV proton colliders such 
as the CERN Large Hadron Collider (LHC), $t{\bar t}$
production will be enormously larger than the Tevatron rates, and  the 
accuracy with which the top mass can be measured in proton colliders 
is around 2~GeV\cite{ATLAS}. This makes the determination of production 
cross section at LHC more accurately than at Fermilab Tevatron. Processes
related to the top quark (the heaviest fermion so far found) may be sensitive
to possible new physics beyond the SM, especially the non-standard Higgs 
sector, due to the large Yukawa coupling.

The simplest extension of the SM Higgs sector is the two-Higgs-doublet models
(2HDM) which can give an enhancement of the Yukawa couplings and may 
result in a considerable correction to the top quark production cross
section.  Yukawa correction to the top quark production in general 2HDM 
and in the minimal supersymmetric model(MSSM)  at 
Fermilab Tevatron  has been carried out in Ref. \cite{STANGE}, where   
large corrections are found.  Supersymmetric electroweak correction 
\cite{YJM} and QCD correction\cite{LCS} to top quark production 
at the Tevatron are also found to be large.   
At LHC the main production mechanism of top quark is the gluon-gluon 
fusion process $gg\rightarrow t {\bar t}$. 
The full electroweak  corrections to this process as well as $qq\rightarrow 
t{\bar t}$ in the Standard Model have been calculated in Ref. \cite{HOLLIK}. 
In this paper we investigate the Yukawa correction to the top quark 
production by the process $gg\rightarrow t{\bar t}$ in a general 2HDM 
and in the minimal supersymmetric model at LHC. 
Our calculation of the corrections to the S-matrix elements will be given in 
Sec.II. In Sec III, we present our numerical results and discussions of the 
cross-sections in the SM , 2HDM and MSSM .  

\begin{flushleft} {\Large\bf II. Calculation of the S-matrix Elements} 
\end{flushleft}

	The tree-level Feynman diagrams and the relevant Yukawa 
corrections to $gg\rightarrow t{\bar t}$ are shown in Fig.1 (u-channel
diagrams of Fig.1b and Fig.1d--Fig.1h are not explicitly shown) in which the 
dashed lines in Fig.1c and Fig.1e--Fig.1h represent $H^0$, $h$, $A$, $G^0$, 
$H^+$ and $G^+$ ($G^0$ and $G^+$ are, respectively, Goldstone bosons of $Z$ 
and $W^+$), while those in Fig.1d represent only  $H^0$ and  $h$. 
The relavant Feynman rules can be found in Ref.\cite{HABER}. 
 
At tree level, the S-matrix element is composed of three different production 
channels(s-,t-,u-channel) as follows:

\begin{eqnarray}                                                %(1)-(3)
 M_0^{s} &= & -ig_s^2if_{abc}T^c_{ji}\Gamma^\mu{\bar u}(p_2)
 \gamma_\mu v(p_1)/\hat{s} \\
 M_0^t& = & -ig_s^2(T^bT^a)_{ji}\epsilon(p_4)^\mu\epsilon(p_3)^\nu{\bar u}(p_2)
 \gamma_\mu(\rlap/p_2-\rlap/p_4+m_t)\gamma_\nu v(p_1)/(\hat{t}-m_t^2)\\
 M_0^u& = & M_0^t(p_3\leftrightarrow p_4,\;\;T^a\leftrightarrow T^b,\;\;
 \hat{t}\rightarrow \hat{u} ),
\end{eqnarray}
where  $\Gamma^\mu$ is given in the Appendix. Instead of 
calculating the square of the amplitudes explicitly,  we calculate  
the amplitudes numerically by using the method of Ref. \cite{ZEPPEN}. 
This method greatly simplifies our calculations.   

The $O(\alpha m_t^2/m_W^2)$ Yukawa corrections to $gg\rightarrow t{\bar t}$ 
are shown in Fig.1c--Fig.1h, where Fig.1d alone is gauge invariant 
and the sum of Fig.1c,Fig.1e--Fig.1h is gauge invariant. In our calculation, 
we use dimensional regularization to regulate the ultaviolet divergences and 
adopt the on-mass-shell renormalization scheme.

We only give the explicit results of the s- and t-channel contributions to 
the total Yukawa correction. The u-channel results can be obtained by the 
following substitutions:
\begin{equation}\begin{array}{l}                               %(4)
p_3\leftrightarrow p_4,\;\;
T^a\leftrightarrow T^b,\;\;
\hat{t}\leftrightarrow\hat{u}.
\end{array}\end{equation}

Fig.1c leads to the s-channel vertex correction $\delta M^{s1}$ :
\begin{equation}\begin{array}{l}                                   %(5)
\displaystyle\delta M^{s1}=\frac{\alpha m_t^2}{16\pi m_W^2s_W^2} 
 (-ig_s^2) if_{abc}T^c_{ji}\Gamma^\mu
{\bar u}(p_2)(f_1^{s1}\gamma_\mu+f_2^{s1}p_{2\mu})v(p_1)/\hat{s} 
\end{array}\end{equation}

The s-channel Higgs exchange of Fig.1d gives $\delta M^{s2,t}$:
\begin{equation}\begin{array}{ll}                                 %(6)
\delta M^{s2,t}= & \displaystyle\frac{\alpha m_t^2}{16\pi m_W^2s_W^2} 
  (-ig_s^2)\frac{\delta_{ab}}{2}\delta_{ji}
\displaystyle\sum\limits_{i=H^0,h}\eta_i\frac{1}{\hat{s}-m_i^2+im_i\Gamma_i}
\displaystyle\epsilon(p_4)^\mu\epsilon(p_3)^\nu\\
& {\bar u}(p_2)(f_1^{s2}g_{\mu\nu}+f_2^{s2}p_{3\mu}p_{4\nu})v(p_1)
\end{array}\end{equation}

The top quark self-energy $\delta M^{self,t}$ of Fig.1e is:  
\begin{equation}\begin{array}{ll}                                %(7)
\delta M^{self,t}= & \displaystyle\frac{\alpha m_t^2}{16\pi m_W^2s_W^2}
  (-ig_s^2)(T^bT^a)_{ji}\epsilon(p_4)^\mu\epsilon(p_3)^\nu
  {\bar u}(p_2)\gamma_\mu(\rlap/p_2-\rlap/p_4+m_t)\\
& [f_1^{self,t}+f_2^{self,t}(\rlap/p_2-\rlap/p_4)]
  (\rlap/p_2-\rlap/p_4+m_t)\gamma_\nu v(p_1)/(\hat{t}-m_t^2)
\end{array}\end{equation}

$\delta M^{v1,t}$ and $\delta M^{v2,t}$ of the vertex corrections 
(Fig.1f--Fig.1g)  are:
\begin{equation}\begin{array}{ll}                                 %(8)
\delta M^{v1,t}= & \displaystyle\frac{\alpha m_t^2}{16\pi m_W^2s_W^2}(-ig_s^2)
(T^bT^a)_{ji}\epsilon(p_4)^\mu\epsilon(p_3)^\nu{\bar u}(p_2)
(f_1^{v1,t}\gamma_\mu+f_2^{v1,t}p_{2\mu}\\
& +f_4^{v1,t}\gamma_\mu \rlap/p_4+f_5^{v1,t}p_{2\mu} \rlap/p_4)
(\rlap/p_2-\rlap/p_4+m_t)\gamma_\nu v(p_1)/(\hat{t}-m_t^2)
\end{array}\end{equation}

\begin{equation}\begin{array}{ll}                                   %(9)
\delta M^{v2,t}= & \displaystyle\frac{\alpha m_t^2}{16\pi m_W^2s_W^2}(-ig_s^2)
(T^bT^a)_{ji}\epsilon(p_4)^\mu\epsilon(p_3)^\nu{\bar u}(p_2)
\gamma_\mu(\rlap/p_2-\rlap/p_4+m_t)\\
& (f_1^{v2,t}\gamma_\nu+f_2^{v2,t}p_{1\nu}+f_4^{v2,t}\rlap/p_3\gamma_\nu
+f_5^{v2,t}\rlap/p_3p_{1\nu}) v(p_1)/(\hat{t}-m_t^2)
\end{array}\end{equation}

$\delta M^{box,t}$ of the box diagram (Fig.1h) is:
\begin{equation}\displaystyle\begin{array}{ll}                      %(10)
\delta M^{box,t}=&
\displaystyle \frac{\alpha m_t^2}{16\pi m_W^2s_W^2}(-ig_s^2)
  (T^bT^a)_{ji}\epsilon(p_4)^\mu
  \epsilon(p_3)^\nu{\bar u}(p_2)
  [f_1^{b,t}\gamma_\nu\gamma_\mu 
  +f_2^{b,t}\gamma_\mu\gamma_\nu\\
& +f_3^{b,t}p_{1\nu}\gamma_\mu
  +f_4^{b,t}p_{1\mu}\gamma_\nu+f_5^{b,t}p_{2\nu}\gamma_\mu
  +f_6^{b,t}p_{2\mu}\gamma_\nu +f_7^{b,t}p_{1\mu}p_{1\nu}\\
& +f_8^{b,t}p_{1\mu}p_{2\nu}+f_9^{b,t}p_{2\mu}p_{1\nu}
  +f_{10}^{b,t}p_{2\mu}p_{2\nu}
  +\rlap/p_4 (f_{11}^{b,t}\gamma_\nu\gamma_\mu  \\
& +f_{12}^{b,t}\gamma_\mu\gamma_\nu+f_{13}^{b,t}p_{1\nu}\gamma_\mu
  +f_{14}^{b,t}p_{1\mu}\gamma_\nu+f_{15}^{b,t}p_{2\nu}\gamma_\mu
  +f_{16}^{b,t}p_{2\mu}\gamma_\nu  \\
& +f_{17}^{b,t}p_{1\mu}p_{1\nu}+f_{18}^{b,t}p_{1\mu}p_{2\nu}
  +f_{19}^{b,t}p_{2\mu}p_{1\nu}+f_{20}^{b,t}p_{2\mu}p_{2\nu})]v(p_1)
\end{array}\end{equation}

The following parameters are used in our calculation:
\begin{equation}\begin{array}{l}                                  %(11)
\sqrt{s}=14 TeV,\;\;m_t=176 GeV,\;\;\\
{\rm MRS~parton~ distribution~ set~ A^\prime}\cite{St},\;\;
Q^2=\hat{s},\\
|\eta|<2.5,\;\;p_T>20\;GeV.\\
\end{array}\end{equation} 
The $\eta$ and $p_T$ cuts are used in order to avoid the singularities 
in the calculation of form factors and increase the relative corrections. 

\begin{flushleft} {\Large\bf III. Numerical Results and Discussions} \end{flushleft}

In our calculation, we checked our program with gauge invariance and 
it is accurate to $10^{-10}$.   For comparison, we first give the 
results of the SM in Fig.2. We see that the SM results 
are always negative and do not exceed $-1.5\%$. The shape of the curve is 
like  that of Ref. \cite{HOLLIK}. 
In Fig.3  and Fig.4 we present the results of the general 2HDM 
and the MSSM with a small $\beta=0.25$ as in Ref. \cite{STANGE}.
In the general 2HDM, the correction can be positive or negative 
depending on the value of $m_H=m_h$. The positive correction can reach 
about $5\%$ and the negative about $-7\%$ which may be
potentially  observable experimentally. 
The MSSM results are all negative and 
can reach as large as $-10\%$ for $m_{H^\pm}<400$ GeV.
The results of the MSSM with $\tan \beta =1$ are given in Fig. 5.     
The corrections never exceed $-2\%$. 
Therefore, if $\tan\beta$ cannot 
be small as may be in the case of the MSSM, the Yukawa corrections 
will be too small to be observable.  Although there is a factor 
$\displaystyle\frac{m_b^2}{m_t^2}\tan^2\beta$ in $\eta_{H^+}$,  
the results  vary insensitively with $\beta$ in the large $\tan\beta$ region 
as shown in Fig.6 for the MSSM with $\tan \beta =70$. 
This conclusion is also valid for a general 2HDM.  

Therefore, we come to the conclusion that for small $\beta$, in 
the general 2HDM and the MSSM, the Yukawa corrections 
to $gg\rightarrow t \bar t $      
at  LHC can  exceed about $-10\%$ which may be potentially observable.  
For larger $\tan\beta =1$, the Yukawa corrections  in the 
general 2HDM and the MSSM will not be observable.  

This work is supported in part by the National Natural Science Foundation of
China, the Fundamental Research Foundation of Tsinghua University and 
a grant from the State Commission of Science and Technology of China.
 
\newpage
\begin{flushleft}
{\Large\bf Appendix }
\end{flushleft}

We give here the form factors for the matrix element appeared in the text.  
They are written in terms of the conventional one-, two-, three- and 
four-point scalar loop integrals defined in Ref.\cite{VELTMAN}.  
We set $M_Z$ and $M_W$ to zero in the numerical calculation as in Ref. 
\cite{STANGE}.

\begin{eqnarray*}
f_1^{s1}&= & -\sum\limits_{i=H^0,h}\eta_i[\frac{1}{2}-2C_{24}-Z_i^v
+m_t^2(4C_0-C_{21})\\
&  &  -k^2(C_{22}-c_{23})](-p_2,k,m_i,m_t,m_t)\\
&  &  -\sum\limits_{i=A,Z}\eta_i[\frac{1}{2}-2C_{24}-Z_i^v-m_t^2C_{21}\\
&  &   -k^2(C_{22}-C_{23})](-p_2,k,m_i,m_t,m_t)\\
& &-\sum\limits_{i=H^+,W^+}\eta_i[\frac{1}{2}-2C_{24}-Z_i^v+m_t^2(C_0-C_{21})\\
&  & -k^2(C_{22}-C_{23})](-p_2,k,m_i,m_b,m_b)\\
f_2^{s1}&= & -\sum\limits_{i=H^0,h}\eta_i[2m_t(2C_{11}+C_{21})](-p_2,k,m_i,
m_t,m_t)\\
&  & -\sum\limits_{i=A,Z}\eta_i[2m_tC_{21}](-p_2,k,m_i,m_t,m_t)\\
&  & -\sum\limits_{i=H^+,W^+}\eta_i[2m_t(C_{11}+C_{21})](-p_2,k,m_i,m_b,m_b)\\
f_1^{s2,t}&= & 4m_t[\frac{1}{2}+m_t^2C_0-p_3\cdot p_4(2C_{22}-2C_{23}+C_0)]
(p_4,-k,m_t,m_t,m_t)\\
f_2^{s2,t}&= & 4m_t[C_0+4(C_{22}-C_{23})](p_4,-k,m_t,m_t,m_t)\\
f_1^{self,t}&= & \left\{\sum\limits_{i=H^0,h}\eta_i [m_t(-B_0
+Z^v_i+\delta m_i)](\hat{t},m_t,m_i)\right.\\
 & &\sum\limits_{i=A,Z}\eta_i [m_t(B_0+Z^v_i+\delta m_i)](\hat{t},m_t,m_i)\\
 & &\left.\sum\limits_{i=H^+,W^+}\eta_i [m_t(Z^v_i+\delta m_i)](\hat{t},m_b,m_i)
 \right\}/(\hat{t}-m_t^2) \\
f_2^{self,t}&= & \left\{\sum\limits_{i=H^0,h}\eta_i [B_1-Z^v_i]
(\hat{t},m_t,m_i)\right.\\
 & &\sum\limits_{i=A,Z}\eta_i [B_1-Z^v_i](\hat{t},m_t,m_i)\\
 & &\left.\sum\limits_{i=H^+,W^+}\eta_i [B_1-Z^v_i](\hat{t},m_b,m_i)\right\}
 /(\hat{t}-m_t^2) \\
f_1^{v1,t}&= & -\sum\limits_{i=H^0,h}\eta_i[\frac{1}{2}-2C_{24}-Z_i^v
    -m_t^2(2C_{11}+C_{21})\\
 & & +2p_2\cdot p_4(C_{12}+C_{23})](-p_2,p_4,m_i,m_t,m_t)\\
 & & -\sum\limits_{i=A,Z}\eta_i[\frac{1}{2}-2C_{24}-Z_i^v-m_t^2(2C_{11}
    +C_{21})\\
 & & +2p_2\cdot p_4(C_{12}+C_{23})](-p_2,p_4,m_i,m_t,m_t)\\
 & & -\sum\limits_{i=H^+,W^+}\eta_i[\frac{1}{2}-2C_{24}-Z_i^v-m_t^2(2C_{11}
    +C_{21}+C_0)\\
 & & +2p_2\cdot p_4(C_{12}+C_{23})](-p_2,p_4,m_i,m_b,m_b)\\
f_2^{v1,t}&= & -\sum\limits_{i=H^0,h}\eta_i[2m_t(3C_{11}+C_{21}+2C_0)]
    (-p_2,p_4,m_i,m_t,m_t)\\
& &-\sum\limits_{i=A,Z}\eta_i[2m_t(C_{11}+C_{21})](-p_2,p_4,m_i,m_t,m_t)\\
& &-\sum\limits_{i=H^+,W^+}\eta_i[2m_t(2C_{11}+C_{21}+C_0)]
    (-p_2,p_4,m_i,m_b,m_b)\\
f_4^{v1,t}&= & \sum\limits_{i=H^0,h}\eta_i[m_t(2C_0+C_{11})]
    (-p_2,p_4,m_i,m_t,m_t)\\
& &\sum\limits_{i=A,Z}\eta_i[m_tC_{11}](-p_2,p_4,m_i,m_t,m_t)\\
& &\sum\limits_{i=H^+,W^+}\eta_i[m_t(C_{11}+C_0)](-p_2,p_4,m_i,m_b,m_b)\\
f_5^{v1,t}&= & \sum\limits_{i=H^0,h}\eta_i[2(C_{12}+C_{23})]
    (-p_2,p_4,m_i,m_t,m_t)\\
& &\sum\limits_{i=A,Z}\eta_i[2(C_{12}+C_{23})](-p_2,p_4,m_i,m_t,m_t)\\
& &\sum\limits_{i=H^+,W^+}\eta_i[2(C_{12}+C_{23})](-p_2,p_4,m_i,m_b,m_b)\\
f_1^{v2,t}&= & -\sum\limits_{i=H^0,h}\eta_i[\frac{1}{2}-2C_{24}-Z_i^v
    -m_t^2(2C_{11}+C_{21})\\
 & & +2p_1\cdot p_3(C_{12}+C_{23})](p_1,-p_3,m_i,m_t,m_t)\\
 & & -\sum\limits_{i=A,Z}\eta_i[\frac{1}{2}-2C_{24}-Z_i^v-m_t^2(2C_{11}
    +C_{21})\\
 & &+2p_1\cdot p_3(C_{12}+C_{23})](p_1,-p_3,m_i,m_t,m_t)\\
 & & -\sum\limits_{i=H^+,W^+}\eta_i[\frac{1}{2}-2C_{24}-Z_i^v-m_t^2(2C_{11}
    +C_{21}+C_0)\\
 & & +2p_1\cdot p_3(C_{12}+C_{23})](p_1,-p_3,m_i,m_b,m_b)\\
f_2^{v2,t}&= & \sum\limits_{i=H^0,h}\eta_i[2m_t(3C_{11}+C_{21}+2C_0)]
    (p_1,-p_3,m_i,m_t,m_t)\\
& &\sum\limits_{i=A,Z}\eta_i[2m_t(C_{11}+C_{21})](p_1,-p_3,m_i,m_t,m_t)\\
& &\sum\limits_{i=H^+,W^+}\eta_i[2m_t(2C_{11}+C_{21}+C_0)]
    (p_1,-p_3,m_i,m_b,m_b)\\
f_4^{v2,t}&= -& \sum\limits_{i=H^0,h}\eta_i[m_t(2C_0+C_{11})]
    (p_1,-p_3,m_i,m_t,m_t)\\
& &-\sum\limits_{i=A,Z}\eta_i[m_tC_{11}](p_1,-p_3,m_i,m_t,m_t)\\
& &-\sum\limits_{i=H^+,W^+}\eta_i[m_t(C_{11}+C_0)](p_1,-p_3,m_i,m_b,m_b)\\
f_5^{v2,t}&= & \sum\limits_{i=H^0,h}\eta_i[2(C_{12}+C_{23})]
    (p_1,-p_3,m_i,m_t,m_t)\\
& &\sum\limits_{i=A,Z}\eta_i[2(C_{12}+C_{23})](p_1,-p_3,m_i,m_t,m_t)\\
& &\sum\limits_{i=H^+,W^+}\eta_i[2(C_{12}+C_{23})](p_1,-p_3,m_i,m_b,m_b)
\end{eqnarray*}    %
\begin{eqnarray*}
f_1^{b,t}&= & -\sum\limits_{i=H^0,h}\eta_i[f_1^{q2q3}+4m_tD_{27}](-p_2,p_4,
   p_3,m_i,m_t,m_t,m_t)\\
& &-\sum\limits_{i=A,Z}\eta_i[f_1^{q2q3}](-p_2,p_4,p_3,m_i,m_t,m_t,m_t)\\
& &-\sum\limits_{i=H^+,W^+}\eta_i[f_1^{q2q3}+2m_tD_{27}]
   (-p_2,p_4,p_3,m_i,m_b,m_b,m_b)\\
f_2^{b,t}&= & -\sum\limits_{i=H^0,h}\eta_i[f_2^{q2q3}+3m_t(-m_t^2D_{21}
    +2p_2\cdot p_4D_{24}\\
& & +2p_2\cdot p_3D_{25}-2p_3\cdot p_4D_{26})-8m_tD_{27}\\
& & +4m_t(-m_t^2D_{11}
+p_2\cdot p_4(D_{11}+D_{12}-D_{13}))](-p_2,p_4,p_3,m_i,m_t,m_t,m_t)\\
& &-\sum\limits_{i=A,Z}\eta_i[f_2^{q2q3}+m_t(-m_t^2D_{21}
  +2p_2\cdot p_4D_{24}\\
& &+2p_2\cdot p_3D_{25}-2p_3\cdot p_4D_{26})-4m_tD_{27}]
(-p_2,p_4,p_3,m_i,m_t,m_t,m_t)\\
& &-\sum\limits_{i=H^+,W^+}\eta_i[f_2^{q2q3}+2m_t(-m_t^2D_{21}  
  +2p_2\cdot p_4D_{24}\\
& & +2p_2\cdot p_3D_{25}-2p_3\cdot p_4D_{26})-6m_tD_{27}+m_t(-3m_t^2D_{11}\\
& &  +2p_2\cdot p_4(D_{11}+D_{12}-D_{13}))-m_t^3D_{0}]
(-p_2,p_4,p_3,m_i,m_b,m_b,m_b)\\
f_3^{b,t}&= &-\sum\limits_{i=H^0,h}\eta_i[f_3^{q2q3}]
(-p_2,p_4,p_3,m_i,m_t,m_t,m_t)\\
& &-\sum\limits_{i=A,Z}\eta_i[f_3^{q2q3}]
(-p_2,p_4,p_3,m_i,m_t,m_t,m_t)\\
& &-\sum\limits_{i=H^+,W^+}\eta_i[f_3^{q2q3}+2m_t^2(D_{12}-D_{13})]
(-p_2,p_4,p_3,m_i,m_b,m_b,m_b)\\
f_4^{b,t}&= &-\sum\limits_{i=H^0,h}\eta_i[f_4^{q2q3}-8m_t^2D_{13}]
(-p_2,p_4,p_3,m_i,m_t,m_t,m_t)\\
& &-\sum\limits_{i=A,Z}\eta_i[f_4^{q2q3}]
(-p_2,p_4,p_3,m_i,m_t,m_t,m_t)\\
& &-\sum\limits_{i=H^+,W^+}\eta_i[f_4^{q2q3}-2m_t^2D_{13}]
(-p_2,p_4,p_3,m_i,m_b,m_b,m_b)\\
f_5^{b,t}&= &-\sum\limits_{i=H^0,h}\eta_i[f_5^{q2q3}]
(-p_2,p_4,p_3,m_i,m_t,m_t,m_t)\\
& &-\sum\limits_{i=A,Z}\eta_i[f_5^{q2q3}]
(-p_2,p_4,p_3,m_i,m_t,m_t,m_t)\\
& &-\sum\limits_{i=H^+,W^+}\eta_i[f_5^{q2q3}+2m_t^2(D_{12}-D_{11})]
(-p_2,p_4,p_3,m_i,m_b,m_b,m_b)\\
f_6^{b,t}&= &-\sum\limits_{i=H^0,h}\eta_i[f_6^{q2q3}+8m_t^2(D_{0}
+D_{11}-D_{13})](-p_2,p_4,p_3,m_i,m_t,m_t,m_t)\\
& &-\sum\limits_{i=A,Z}\eta_i[f_6^{q2q3}]
(-p_2,p_4,p_3,m_i,m_t,m_t,m_t)\\
& &-\sum\limits_{i=H^+,W^+}\eta_i[f_6^{q2q3}+2m_t^2(D_0+2D_{11}-2D_{13})]
(-p_2,p_4,p_3,m_i,m_b,m_b,m_b)\\
f_7^{b,t}&= &-\sum\limits_{i=H^0,h}\eta_i[f_7^{q2q3}+8m_tD_{26}]
(-p_2,p_4,p_3,m_i,m_t,m_t,m_t)\\
& &-\sum\limits_{i=A,Z}\eta_i[f_7^{q2q3}]
(-p_2,p_4,p_3,m_i,m_t,m_t,m_t)\\
& &-\sum\limits_{i=H^+,W^+}\eta_i[f_7^{q2q3}+4m_tD_{26}]
(-p_2,p_4,p_3,m_i,m_b,m_b,m_b)\\
f_8^{b,t}&= &-\sum\limits_{i=H^0,h}\eta_i[f_8^{q2q3}+8m_t(D_{26}-D_{25})]
(-p_2,p_4,p_3,m_i,m_t,m_t,m_t)\\
& &-\sum\limits_{i=A,Z}\eta_i[f_8^{q2q3}]
(-p_2,p_4,p_3,m_i,m_t,m_t,m_t)\\
& &-\sum\limits_{i=H^+,W^+}\eta_i[f_8^{q2q3}+4m_t(D_{26}-D_{25})]
(-p_2,p_4,p_3,m_i,m_b,m_b,m_b)\\
f_9^{b,t}&= &-\sum\limits_{i=H^0,h}\eta_i[f_9^{q2q3}
+8m_t(D_{26}-D_{24}-D_{12})](-p_2,p_4,p_3,m_i,m_t,m_t,m_t)\\
& &-\sum\limits_{i=A,Z}\eta_i[f_9^{q2q3}]
(-p_2,p_4,p_3,m_i,m_t,m_t,m_t)\\
& &-\sum\limits_{i=H^+,W^+}\eta_i[f_9^{q2q3}+4m_t(D_{26}-D_{24}-D_{12})]
(-p_2,p_4,p_3,m_i,m_b,m_b,m_b)\\
f_{10}^{b,t}&= &-\sum\limits_{i=H^0,h}\eta_i[f_{10}^{q2q3}+8m_t(D_{21}+D_{26}\\
& & -D_{24}-D_{25}+D_{11}-D_{12})](-p_2,p_4,p_3,m_i,m_t,m_t,m_t)\\
& &-\sum\limits_{i=A,Z}\eta_i[f_{10}^{q2q3}](-p_2,p_4,p_3,m_i,m_t,m_t,m_t)\\
& &-\sum\limits_{i=H^+,W^+}\eta_i[f_{10}^{q2q3}+4m_t(D_{21}+D_{26}\\
& & -D_{24}-D_{25}+D_{11}-D_{12})](-p_2,p_4,p_3,m_i,m_b,m_b,m_b)\\
f_{11}^{b,t}&= &-\sum\limits_{i=H^0,h}\eta_i[f_{11}^{q2q3}]
           (-p_2,p_4,p_3,m_i,m_t,m_t,m_t)\\
& &-\sum\limits_{i=A,Z}\eta_i[f_{11}^{q2q3}]
           (-p_2,p_4,p_3,m_i,m_t,m_t,m_t)\\
& &-\sum\limits_{i=H^+,W^+}\eta_i[f_{11}^{q2q3}]
(-p_2,p_4,p_3,m_i,m_b,m_b,m_b)\\
f_{12}^{b,t}&= &-\sum\limits_{i=H^0,h}\eta_i[f_{12}^{q2q3}+4m_t^2D_0]
(-p_2,p_4,p_3,m_i,m_t,m_t,m_t)\\
& &-\sum\limits_{i=A,Z}\eta_i[f_{12}^{q2q3}]
(-p_2,p_4,p_3,m_i,m_t,m_t,m_t)\\
& &-\sum\limits_{i=H^+,W^+}\eta_i[f_{12}^{q2q3}+m_t^2(D_0+D_{12}-D_{13})]
(-p_2,p_4,p_3,m_i,m_b,m_b,m_b)\\
f_{13}^{b,t}&= &-\sum\limits_{i=H^0,h}\eta_i[f_{13}^{q2q3}-4m_tD_{12}]
(-p_2,p_4,p_3,m_i,m_t,m_t,m_t)\\
& &-\sum\limits_{i=A,Z}\eta_i[f_{13}^{q2q3}]
(-p_2,p_4,p_3,m_i,m_t,m_t,m_t)\\
& &-\sum\limits_{i=H^+,W^+}\eta_i[f_{13}^{q2q3}-2m_tD_{12}]
(-p_2,p_4,p_3,m_i,m_b,m_b,m_b)\\
f_{14}^{b,t}&= &-\sum\limits_{i=H^0,h}\eta_i[f_{14}^{q2q3}+4m_tD_{13}]
(-p_2,p_4,p_3,m_i,m_t,m_t,m_t)\\
& &-\sum\limits_{i=A,Z}\eta_i[f_{14}^{q2q3}]
(-p_2,p_4,p_3,m_i,m_t,m_t,m_t)\\
& &-\sum\limits_{i=H^+,W^+}\eta_i[f_{14}^{q2q3}+2m_tD_{13}]
(-p_2,p_4,p_3,m_i,m_b,m_b,m_b)\\
f_{15}^{b,t}&= &-\sum\limits_{i=H^0,h}\eta_i[f_{15}^{q2q3}+4m_t(D_{11}-D_{12})]
(-p_2,p_4,p_3,m_i,m_t,m_t,m_t)\\
& &-\sum\limits_{i=A,Z}\eta_i[f_{15}^{q2q3}]
(-p_2,p_4,p_3,m_i,m_t,m_t,m_t)\\
& &-\sum\limits_{i=H^+,W^+}\eta_i[f_{15}^{q2q3}+2m_t(D_{11}-D_{12})]
(-p_2,p_4,p_3,m_i,m_b,m_b,m_b)\\
f_{16}^{b,t}&= &-\sum\limits_{i=H^0,h}\eta_i[f_{16}^{q2q3}+4m_t(D_{13}-D_{11})]
(-p_2,p_4,p_3,m_i,m_t,m_t,m_t)\\
& &-\sum\limits_{i=A,Z}\eta_i[f_{16}^{q2q3}]
(-p_2,p_4,p_3,m_i,m_t,m_t,m_t)\\
& &-\sum\limits_{i=H^+,W^+}\eta_i[f_{16}^{q2q3}+2m_t(D_{13}-D_{11})]
(-p_2,p_4,p_3,m_i,m_b,m_b,m_b)\\
f_{17}^{b,t}&= &-\sum\limits_{i=H^0,h}\eta_i[f_{17}^{q2q3}]
(-p_2,p_4,p_3,m_i,m_t,m_t,m_t)\\
& &-\sum\limits_{i=A,Z}\eta_i[f_{17}^{q2q3}]
(-p_2,p_4,p_3,m_i,m_t,m_t,m_t)\\
& &-\sum\limits_{i=H^+,W^+}\eta_i[f_{17}^{q2q3}]
(-p_2,p_4,p_3,m_i,m_b,m_b,m_b)\\
f_{18}^{b,t}&= &-\sum\limits_{i=H^0,h}\eta_i[f_{18}^{q2q3}]
(-p_2,p_4,p_3,m_i,m_t,m_t,m_t)\\
& &-\sum\limits_{i=A,Z}\eta_i[f_{18}^{q2q3}]
(-p_2,p_4,p_3,m_i,m_t,m_t,m_t)\\
& &-\sum\limits_{i=H^+,W^+}\eta_i[f_{18}^{q2q3}]
(-p_2,p_4,p_3,m_i,m_b,m_b,m_b)\\
f_{19}^{b,t}&= &-\sum\limits_{i=H^0,h}\eta_i[f_{19}^{q2q3}]
(-p_2,p_4,p_3,m_i,m_t,m_t,m_t)\\
& &-\sum\limits_{i=A,Z}\eta_i[f_{19}^{q2q3}]
(-p_2,p_4,p_3,m_i,m_t,m_t,m_t)\\
& &-\sum\limits_{i=H^+,W^+}\eta_i[f_{19}^{q2q3}]
(-p_2,p_4,p_3,m_i,m_b,m_b,m_b)\\
f_{20}^{b,t}&= &-\sum\limits_{i=H^0,h}\eta_i[f_{20}^{q2q3}]
(-p_2,p_4,p_3,m_i,m_t,m_t,m_t)\\
& &-\sum\limits_{i=A,Z}\eta_i[f_{20}^{q2q3}]
(-p_2,p_4,p_3,m_i,m_t,m_t,m_t)\\
& &-\sum\limits_{i=H^+,W^+}\eta_i[f_{20}^{q2q3}]
(-p_2,p_4,p_3,m_i,m_b,m_b,m_b)\\
  & & \\
f_1^{q2q3}&= & 2m_tD_{311}\\
f_2^{q2q3}&= & -m_t^3D_{31}+2m_t (p_2\cdot p_4D_{34}+p_2\cdot p_3D_{35}
-p_3\cdot p_4D_{310}-2D_{311})\\
 & & -m_t^3D_{21}+2m_t[p_2\cdot p_4(D_{21}-D_{25})-p_2\cdot p_3D_{25}
+p_3\cdot p_4 D_{26}+D_{27}]\\
f_3^{q2q3}&= &  2[m_t^2(D_{34}-D_{35})+2p_2\cdot p_3(D_{37}-D_{310})
+2p_3\cdot p_4(D_{38}-D_{39})\\
 & & +2p_2\cdot p_4(D_{310}-D_{36})+4D_{312}
-6D_{313}]\\
 & & +4m_t^2(D_{24}-D_{25})+4p_2\cdot p_4(D_{25}+D_{26}-D_{22}-D_{23})
+4D_{27}\\
f_4^{q2q3}&= & 2[m_t^2D_{35}-2p_2\cdot p_3D_{37}
+2p_3\cdot p_4D_{39}-2p_2\cdot p_4D_{310}+4D_{313}]\\
 & &+4p_2\cdot p_4(D_{23}-D_{25})\\
f_5^{q2q3}&= & 2[m_t^2(D_{34}-D_{31})+2p_2\cdot p_4(D_{34}-D_{36})
+2p_2\cdot p_3(D_{35}-D_{310})\\
 & &+2p_3\cdot p_4(D_{38}-D_{310})
-4D_{311}+4D_{312}]\\
 & &+4m_t^2(D_{24}-D_{21})+4p_2\cdot p_4(D_{24}-D_{22})\\
f_6^{q2q3}&= & 4(D_{311}-D_{313})+2m_t^2(D_{21}-2D_{25})+4p_2\cdot p_3
D_{25}+4p_2\cdot p_4D_{26}-4p_3\cdot p_4D_{26}\\
f_7^{q2q3}&= & 4m_tD_{310}\\
f_8^{q2q3}&= & 4m_t(D_{310}-D_{35})\\
f_9^{q2q3}&= & 4m_t(D_{310}-D_{34})-4m_tD_{24}\\
f_{10}^{q2q3}&= & 4m_t(D_{31}-D_{34}-D_{35}+D_{310})+4m_t(D_{21}-D_{24})\\
f_{11}^{q2q3}&= & 2(D_{313}-D_{312})-2D_{27}\\
f_{12}^{q2q3}&= & m_t^2(D_{34}-D_{35})+2[p_2\cdot p_4(D_{310}-D_{36})
+p_2\cdot p_3(D_{37}-D_{310})\\
 & & +p_3\cdot p_4(D_{38}-D_{39})+2D_{312}-2D_{313}]\\
 & & +m_t^2(2D_{24}-2D_{25}-D_{21})\\
 & & +2[p_2\cdot p_4(D_{25}+2D_{26}
    -D_{22}-D_{23})+p_2\cdot p_3D_{25}-p_3\cdot p_4 D_{26}]\\
f_{13}^{q2q3}&= & -2m_tD_{24}\\
f_{14}^{q2q3}&= & 2m_tD_{25}\\
f_{15}^{q2q3}&= & 2m_t(D_{21}-D_{24})\\
f_{16}^{q2q3}&= & 2m_t(D_{25}-D_{21})\\
f_{17}^{q2q3}&= & 4(D_{39}-D_{38})+4(D_{23}-D_{26})\\
f_{18}^{q2q3}&= & 4(D_{39}+D_{310}-D_{37}-D_{38})+4(D_{25}-D_{26})\\
f_{19}^{q2q3}&= & 4(D_{36}+D_{39}-D_{38}-D_{310})+4(D_{22}+D_{23}
-D_{25}-D_{26})\\
f_{20}^{q2q3}&= & 4(D_{35}+D_{36}-D_{34}-D_{37}-D_{38}+D_{39})
+4(D_{22}-D_{24}+D_{25}-D_{26}).\\
\end{eqnarray*} 

In the above formulae 
$$\begin{array}{l}
k=p_1+p_2=p_3+p_4,\;\;\hat{s}=k^2,\;\;
\hat{t}=(p_2-p_4)^2,\;\;\hat{u}=(p_2-p_3)^2,\\
\Gamma^\mu=(-p_4+p_3)^\mu\epsilon(p_3)\cdot\epsilon(p_4)+(2p_4+p_3)
\cdot\epsilon(p_3)
\epsilon(p_4)^\mu-(2p_3+p_4)\cdot\epsilon(p_4)\epsilon(p_3)^\mu,\\ 
\displaystyle\eta_{H^0}=\frac{\sin^2\alpha}{\sin^2\beta},\;\; 
\displaystyle\eta_{H^0}=\frac{\cos^2\alpha}{\sin^2\beta},\\
\eta_A=\cot^2\beta,\;\;
\displaystyle\eta_{H^+}=(\cot^2\beta+\frac{m_b^2}{m_t^2}\tan^2\beta),\;\;
\eta_Z=\eta_{W^+}=1.\\
\end{array} $$

The renormalization constants are:\\
$$\displaystyle\begin{array}{l}
Z^v_i=[B_1+2m_t^2(G_0+G_1)](m_t^2,m_t,m_i)\;\; for\;\; i=H^0,\; h\\
Z^v_i=[B_1+2m_t^2(G_1-G_0)](m_t^2,m_t,m_i)\;\; for\;\; i=A,\; Z\\
Z^v_i=[B_1+2m_t^2G_1](m_t^2,m_b,m_i)\;\; for\;\; i=H^+,\; W^+\\
 \mbox{}\\
\delta m_i=[B_0-B_1](m_t^2,m_t,m_i)\;\; for\;\; i=H^0,\; h\\
\delta m_i=[-B_0-B_1](m_t^2,m_t,m_i)\;\; for\;\; i=A,\; Z\\
\delta m_i=[-B_1](m_t^2,m_b,m_i)\;\; for\;\; i=H^+,\; W^+\\
\end{array} $$

where  $\displaystyle G_0=-\frac{dB_0(p^2,m_1,m_2)}{dp^2}|_{p^2=m_t^2}$, 
$\displaystyle G_1=\frac{dB_1(p^2,m_1,m_2)}{dp^2}|_{p^2=m_t^2}$ .

\newpage

\begin{flushleft}
{\Large\bf  Figure Captions}
\end{flushleft}

\parindent=0pt

Fig.1  Feynaman diagrams of tree level and $O(\alpha m_t^2/m_W^2)$ 
Yukawa corrections of $gg\rightarrow t\bar t $.

Fig.2 Results of Yukawa corrections in the Standard Model.

Fig.3 Results of Yukawa corrections in the general Two-Higgs-Doublet 
Model with $\beta=0.25$.

Fig.4 Results of Yukawa corrections in the Minimal Supersymmetric  
Standard Model with $\beta=0.25$.

Fig.5 Results of Yukawa corrections in the Minimal Supersymmetric  
Standard Model with $\beta=\pi/4$.

Fig.6 Results of Yukawa corrections in the Minimal Supersymmetric  
Standard Model with $\tan\beta=70$.

%\newpage
%\mbox{}

%\vspace{16cm}
%\special{BMF=/SSS/ZHY/CAL/GGTTF.BMF}
%\begin{center} {\Large Fig.1} \end{center}
%\newpage
%\mbox{}

%\vspace{16cm}
%\special{BMF=/SSS/ZHY/CAL/re1.BMF}
%\begin{center} {\Large Fig.2} \end{center}
%\newpage
%\mbox{}

%\vspace{16cm}
%\special{BMF=/SSS/ZHY/CAL/re2.BMF}
%\begin{center} {\Large Fig.3} \end{center}
%\newpage
%\mbox{}

%\vspace{16cm}
%\special{BMF=/SSS/ZHY/CAL/re3.BMF}
%\begin{center} {\Large Fig.4} \end{center}
%\newpage
%\mbox{}

%\vspace{16cm}
%\special{BMF=/SSS/ZHY/CAL/re31.BMF}
%\begin{center} {\Large Fig.5} \end{center}

%\newpage
%\mbox{}

%\vspace{16cm}
%\special{BMF=/SSS/ZHY/CAL/re32.BMF}
%\begin{center} {\Large Fig.6} \end{center}


\newpage
\begin{thebibliography}{7}


\bibitem{CDFD0} CDF Collaboration, F. Abe {\it et al.}, Phys. Rev. Lett.
 {\bf 74}, 
2626(1995); D0 Collaboration, S. Abachi {\it et al.}, Phys. Rev. Lett. 
{\bf 74}, 
2632(1995).  

\bibitem{ATLAS} ATLAS Collaboration, CERN/LHCC/94-43(1994)

\bibitem{STANGE} A. Stange and S. Willenbrock, Phys. Rev. {\bf D 48},
 2054(1993).    

\bibitem{HOLLIK} W. Beennakker {\it et al.} Nucl. Phys. {\bf B411}, 343(1994). 

\bibitem{YJM} J.M.Yang and C.S.Li, Phys. Rev. {\bf D 52},1541(1995).
 
\bibitem{LCS} C.S. Li,B.Q. Hu, J.M. Yang and C.G. Hu, Phys. Rev. {\bf D 52},
5014(1995); C.S. Li, H.Y. Zhou, Y.L. Zhu and J.M. Yang, PKU-TH-96-1(1996).  

\bibitem{HABER} H.E. Haber and C.L.Kane, Phys. Rep. 117, 75(1985);
J.F.Gunion and H.E.Haber , Nucl. Phys. {\bf B272},1(1986).

\bibitem{ZEPPEN} K.Hagiwara and D.Zeppenfeld, Nucl. Phys. 
{\bf B313}, 560 (1989); V.Barger, T.Han, and D.Zeppenfeld, Phys.Rev.
{\bf D41},2782(1990).

\bibitem{St} A.D. Martin, W.J. Stirling  and R.G. Roberts,
     Phys. Lett. {\bf B354},155(1995).

\bibitem{VELTMAN} G. Passarino and M.Veltman, Nucl. Phys. {\bf B160}, 
151(1979).

\end{thebibliography}
\end{document}